# An Efficient On-the-Fly Nonadiabatic Coupling Framework Integrated into CP2K

Xiaoke He[1,†], Hsiao-Yi Tsai[1,†], Ziwei Chai[2], JunWen Yin[3,*], Zhongfei Xu[4,*], and Li-Min Liu[1,*]

1 School of Physics, Beihang University, Beijing 100191, China

2 School of Physical Science, Great Bay University, Dongguan, 523000, China

3 Scientific Computing Department, Science and Technology Facilities Council, UK Research and Innovation, Daresbury Laboratory, Keckwick Lane, Daresbury WA4 4AD, United Kingdom

4 College of Environmental Science and Engineering, North China Electric Power University, Beijing 102206, China

*Corresponding authors:

Li-Min Liu, Email: liminliu@buaa.edu.cn; Zhongfei Xu, Email: xuzhongfei@ncepu.edu.cn; JunWen Yin, Email: junwen.yin@stfc.ac.uk

[†]These authors contributed equally to this work

**Abstract**

Nonadiabatic molecular dynamics (NAMD) is widely used to describe hot electron relaxation and nonradiative recombination processes, but high computational costs limit its application to large supercells. Here, we implement a nonadiabatic coupling (NAC) module directly into CP2K, enabling on-the-fly NAC calculation during ab-initio molecular dynamics. Unlike conventional approaches relying on interfaces between external NAMD programs and electronic structure codes, this integration streamlines the workflow. Combined with CP2K's inherent Gaussian and Plane Waves (GPW) method, it achieves efficient NAC calculations for large-scale systems. To ensure numerical stability, a phase correction scheme is introduced to remove inconsistencies of wavefunction phases. Benchmark calculations of hot electron relaxation in crystalline pentacene agree with previous studies, while the results of simulations with and without phase correction show significant difference, confirming the necessity of the phase correction. Simulations on larger pentacene supercells further demonstrate the capability of this implementation for large-scale NAMD simulations.

## 1 Introduction

The nonradiative recombination of charge carriers and hot electron relaxation play important roles in the performance of light conversion materials. Understanding these processes requires describing both atomic-scale structural evolution and the complex electronic structure of materials. Born-Oppenheimer (BO) approximation is employed under equilibrium or quasi-equilibrium conditions, assuming that electrons instantaneously adapt to nuclear motion, while the atomic nuclei are described as evolving on a single adiabatic potential energy surface[1]. When materials undergo photoexcitation or charge transfer, the BO approximation becomes invalid. Under such conditions, the coupling between electrons and nuclei becomes non-negligible, and the effect of nuclear motion on electronic structures needs to be incorporated. The description of nuclear motion based on a single adiabatic potential energy surface no longer sufficient[2]. Theoretical modeling of these processes requires an explicit description of the coupled dynamics of electrons and nuclei, which is commonly addressed within the framework of nonadiabatic molecular dynamics (NAMD)[3,4].

Ideally, a rigorous description of nonadiabatic processes requires a fully quantum mechanical treatment of both electrons and nuclei. However, the computational cost of such an approach increases rapidly with system size, which restricts its applicability to systems containing only a few particles. To overcome this limitation, a variety of mixed quantum-classical (MQC) methods[5] have been developed, such as Ehrenfest mean-field (EMF)[6], trajectory surface hopping (TSH)[7,8], and quantum trajectory mean-field (QTMF)[9]. In these approaches, the electronic subsystem is treated quantum mechanically, while nuclear motion is described using classical trajectories[10]. Unlike mean-field approaches that rely on a single effective surface, TSH constrains nuclei to evolve on the adiabatic potential energy surface at a time while permitting transitions between electronic states. The fewest switches surface hopping (FSSH) algorithm describes nonadiabatic dynamics by propagating classical nuclear trajectories on active adiabatic potential energy surfaces, where stochastic transitions between electronic states are governed by nonadiabatic coupling (NAC), and nuclear motion is affected by electronic transitions through velocity rescaling after surface hops[11,12]. To reduce the computational overhead associated with this coupled dynamics, the classical path approximation (CPA) is introduced. In the CPA-FSSH method, nuclear trajectories are driven on a single reference potential energy surface, neglecting the electronic back-reaction on nuclear motion[13]. This approach has become widely used in surface hopping simulations owing to its computational efficiency. A more detailed description of the CPA-FSSH algorithm is provided in the Supporting Information.

Currently, several computational packages are available for performing NAMD simulations, including PYthon eXtension for Ab Initio Dynamics (PYXAID)[13,14], Hefei-NAMD[15], Libra[16], Newton-X[17,18], and so on. Among these, PYXAID and Hefei-NAMD are often interfaced with widely used first-principles calculation platforms such as Vienna Ab initio Simulation Package (VASP)[19] and Quantum ESPRESSO[20]. While these plane-wave-based methods offer high accuracy, CP2K[21] presents distinct advantages for NAMD simulations of massive supercells owing to its efficient Gaussian and plane wave (GPW) method. By representing the KS matrices with localized Gaussian basis sets while efficiently solving the Hartree potential using plane waves on multi-grids, the GPW approach achieves an optimal balance of accuracy and computational efficiency[22]. Coupled with its excellent parallel scalability and the orbital transformation (OT) algorithm[23], CP2K provides a robust framework for complex systems. Consequently, several advanced NAMD software packages have been interfaced with CP2K. For example, Libra provides a modular platform that supports large-scale simulations through semiempirical methods such as extended tight-binding (xTB)[24]; and Newton-X provides a flexible interface for nonadiabatic dynamics simulations with various coupling schemes, including orbital derivative (OD), local diabatization (LD) and time-dependent Baeck-An (TDBA) approaches[25]. In these interfaces, the electronic structure is calculated within CP2K, and massive amounts of data, such as wavefunctions and overlap matrices, must be printed and subsequently transferred to external NAMD packages for NAC calculation. This process incurs severe disk I/O bottlenecks, drastically reducing computational efficiency and consuming substantial memory. In this work, we develop a module integrated into the CP2K software package for calculating NAC on-the-fly during ab-initio molecular dynamics (AIMD) simulations. A major computational advantage of our implementation is the decoupling of the I/O footprint from the system size. By performing all NAC evaluations natively in memory rather than exporting massive files to disk, the size of the generated storage file is determined solely by the active space (i.e., the subset of relevant electronic states that dominate the nonadiabatic dynamics), effectively eliminating the severe storage bottleneck typically encountered in large-scale simulations. A phase correction method is introduced to the atomic orbital basis, which substantially enhances the numerical stability of NAC. To validate the reliability of the present implementation, the hot electron relaxation in crystalline pentacene was performed as a benchmark test. The resulting relaxation timescales and dynamical processes agree well with previous studies[13,26]. We also investigated the influence of phase correction on the relaxation processes, confirming its necessity in NAMD simulations. The simulation was further extended to a larger supercell, demonstrating the capability of this implementation to efficiently evaluate NAC in large supercells.

## 2 Theory and Methods

The time evolution of system's wavefunction is governed by the time-dependent Schrödinger equation (TDSE):

$$i\hbar\frac{\partial}{\partial t}\Psi(\mathbf{x},\mathbf{X},t) = \hat{\mathbf{H}}(\mathbf{x},\mathbf{X}(t))\Psi(\mathbf{x},\mathbf{X},t) \tag{1}$$

The total wavefunction can be expanded in terms of adiabatic electronic states,

$$\Psi(\mathbf{x},\mathbf{X},t) = \sum_i a_i\,(t)\Phi_i(\mathbf{x};\mathbf{X}) \tag{2}$$

and the corresponding electronic Hamiltonian matrix is

$$H_{el,ij}(\mathbf{X}(t)) = \left\langle \Phi_i(\mathbf{x};\mathbf{X})\middle|\hat{\mathbf{H}}_{el}[\mathbf{X}(t)]\middle|\Phi_j(\mathbf{x};\mathbf{X})\right\rangle \tag{3}$$

Substituting eq 2 and eq 3 into eq 1 and projecting onto $\Phi_i$ yields

$$i\hbar\frac{da_i}{dt} = \sum_j a_j\,[H_{el,ij} - i\hbar\frac{d\mathbf{X}}{dt}\cdot\mathbf{A}_{ij}] \tag{4}$$

and

$$\mathbf{A}_{ij} = \left\langle \Phi_i\middle|\nabla_R\middle|\Phi_j\right\rangle \tag{5}$$

is the nonadiabatic coupling vector.

To reduce the computational complexity associated with many-electron wavefunctions, the electronic problem is treated within the Kohn-Sham (KS) framework, where the interacting many-electron system is projected onto the single-particle representation. In this approximation, many-electron states are represented by Slater determinants constructed from KS orbitals. The Hamiltonian matrix element is written as

$$H_{ij} = \varepsilon_i \delta_{ij} - i\hbar \frac{d\mathbf{X}}{dt} \langle \psi_i | \nabla_R | \psi_j \rangle \tag{6}$$

where $\psi_i$ is a KS orbital.

In practical calculations, we compute NAC on-the-fly during AIMD simulations, so it is easier to compute time-derivative NAC terms. According to eq 5 and the chain rule, we obtain

$$d_{ij}(t) = \frac{d\mathbf{X}}{dt} \cdot \mathbf{A}_{ij} = \left\langle \Phi_i \middle| \frac{\partial}{\partial t} \middle| \Phi_j \right\rangle \tag{7}$$

Thus, by substituting with KS orbitals, the NAC between many-electron states is reduced to matrix elements between KS orbitals:

$$d_{ij}(t) = \left\langle \psi_i \middle| \frac{\partial}{\partial t} \middle| \psi_j \right\rangle \tag{8}$$

Since the electronic wavefunctions are obtained only at discrete time steps, the time-derivative coupling is evaluated using a finite-difference approximation based on overlaps between wavefunctions at consecutive time steps:

$$d_{ij}(t) \approx \frac{\langle \psi_i(t) | \psi_j(t+dt) \rangle - \langle \psi_i(t+dt) | \psi_j(t) \rangle}{2dt} \tag{9}$$

In CP2K, the KS orbitals are expanded in a non-orthogonal atomic orbital basis, therefore, the overlap matrix element $S_{ij}$ is defined to assist with calculations, it takes the form of

$$S_{ij} = \int \mathrm{d}\mathbf{x} \varphi_i^*(\mathbf{x}) \varphi_j(\mathbf{x}) \tag{10}$$

the overlap between orbitals at consecutive time steps is given by:

$$\langle \psi_i(t) | \psi_j(t+dt) \rangle = \sum_{\mu\nu} c_{\mu i}^*(t) S_{\mu\nu}(t, t+dt) c_{\nu j}(t+dt) \tag{11}$$

where $S_{\mu\nu}(t, t+dt)$ is the overlap matrix between atomic orbitals at different time steps, $\varphi_\mu(\mathbf{x})$ is the atomic orbital basis wavefunction, $c_{\mu i}(t)$ is the wavefunction expansion coefficient. Since the term $S_{\mu\nu}(t, t+dt)$ is not directly computed in DFT and is very computationally intensive, for small nuclear displacements, it can be approximated by the instantaneous overlap matrix:

$$
\begin{aligned}
S_{\mu\nu}(t, t+dt) &= \int \mathrm{d}\mathbf{x}\varphi_{\mu}^{*}(\mathbf{x}, t+dt)\varphi_{\nu}(\mathbf{x}, t+dt) \\
&\approx \int \mathrm{d}\mathbf{x}\varphi_{\mu}^{*}(\mathbf{x}, t+dt)\varphi_{\nu}(\mathbf{x}, t) \\
&\approx \int \mathrm{d}\mathbf{x}\varphi_{\mu}^{*}(\mathbf{x}, t)\varphi_{\nu}(\mathbf{x}, t) = S_{\mu\nu}(t)
\end{aligned} \tag{12}
$$

In AIMD simulations, electronic wavefunctions are obtained by independently solving the KS equations at each nuclear configuration. As a result, each electronic eigenstate is defined as an arbitrary phase factor, leading to phase inconsistencies between wavefunctions at consecutive time steps, which may introduce abrupt changes or strong oscillations to NAC during the time evolution, thereby affecting the propagation of electronic coefficients and the evaluation of hopping probabilities[27].

In CP2K, where Gamma-point sampling leads to real-valued wavefunctions, the phase factor is restricted to $\pm 1$, and thus phase inconsistency manifests only as a sign change. Taking advantage of this property, we adopt a phase correction scheme inspired by methods originally developed for plane-wave representations and generalize it to atomic orbital basis. In the plane-wave formalism, a commonly used phase correction coefficient is given by[27]

$$
f_i = \frac{\langle \psi_i(t) \mid \psi_i(t') \rangle}{|\langle \psi_i(t) \mid \psi_i(t') \rangle|} \tag{13}
$$

By expanding wavefunctions in atomic orbital basis, the inner product is evaluated using the wavefunction coefficients and the overlap matrix, the corresponding coefficient can be expressed as

$$
f_i = sign[\sum_{kl} c_k^{*}(t) c_l(t') S_{kl}(t, t')] \tag{14}
$$

To ensure reliable phase alignment, the time interval between $t$ and $t'$ should be sufficiently small. This phase correction procedure enforces phase consistency of the electronic wavefunctions along the AIMD trajectory and significantly enhances the numerical stability of the computed NACs.

The computational workflow in CP2K is primarily divided into two stages: first, nuclear trajectories are obtained during AIMD simulations, then the calculation of NACs between different electronic states based on the instantaneous nuclear configuration is performed. The specific workflow is illustrated in Figure 1.

The electronic structure of the initial configuration is first calculated using OT algorithm to construct the ground-state electron density and Hamiltonian. Then a full diagonalization of the Kohn-Sham Hamiltonian $\widehat{H}_{KS}$ is executed to obtain the initial wavefunction $\psi_{\mathrm{KS}}$, which is stored as the reference state. Subsequently, the system is propagated to the next step using AIMD, yielding an updated configuration. In each step of molecular dynamics, atomic forces and displacements are computed based on OT algorithm. Based on the converged electron density, the $\widehat{H}_{KS}$ is reconstructed and the $\psi_{\mathrm{KS}}$ corresponding to the current time step is obtained. The OT algorithm minimizes the total energy functional without explicitly diagonalizing the Kohn-Sham Hamiltonian. Since NAC calculations require well-defined eigenstates and eigenvalues, an additional diagonalization step is necessary.

The phase correction is achieved by introducing a reference wavefunction and evaluating the overlap between the current and reference wavefunctions. The phase of each orbital is then aligned according to the sign of the overlap, thereby enforcing phase consistency and ensuring stability of NAC. To ensure the validity of the overlap matrix approximation in phase correction, a set of wavefunctions is stored every N steps. After phase correction, the overlaps between orbitals at consecutive time steps are then evaluated, and NACs are computed using a finite-difference scheme. The wavefunctions at current time step are stored for subsequent NAC calculations. This process is cyclically executed in molecular dynamics simulations until the preset number of steps is reached, thereby achieving stable, consistent, and efficient on-the-fly calculation of NAC.

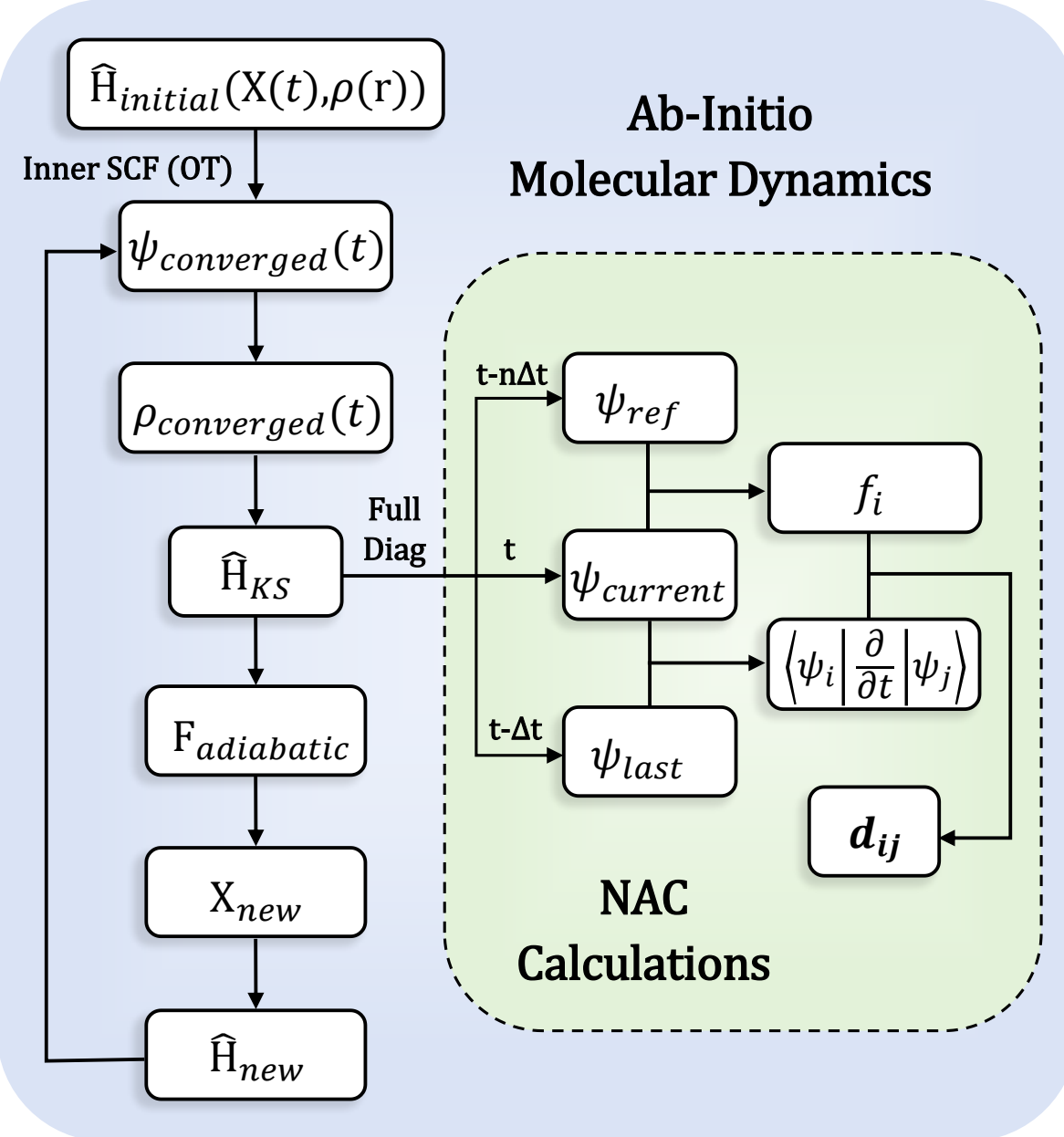

**Figure 1.** Workflow diagram for implementing nonadiabatic coupling matrix calculation in CP2K.

## 3 Study of Hot-electron Relaxation in Pentacene

Pentacene has attracted considerable attention in organic photovoltaics due to its favorable optical absorption, efficient charge transport, and its ability to support singlet fission (SF). In this process, one singlet exciton splits into two triplet excitons, doubling the number of charge carriers and enhancing energy conversion efficiency[28–30]. SF occurs on an ultrafast timescale (within 100 fs)[31] and competes with hot electron relaxation. Therefore, understanding the hot electron relaxation is essential for improving the light conversion efficiency of pentacene.

Pentacene is a layered molecular crystal whose primitive cell contains two molecules arranged in a herringbone pattern, and several polymorphs can be identified depending on the d-spacing perpendicular to the (001) surface[32]. In bulk single crystals, d-spacing is approximately 14.1 Å, and the molecule exhibits a large tilt angle relative to the surface normal[33,34]. In contrast, the monolayer pentacene deposited on a $SiO_2$ substrate exhibits a larger d-spacing of 16.1 Å. Due to the substrate-induced growth, the molecules adopt an orientation that is nearly perpendicular to the (001) surface[35]. Compared with other multilayer thin-film phases, bulk and monolayer pentacene exhibit superior charge transport properties in field-effect transistors[36–38]. The structures of these two phases are shown in Figure 2, and their lattice parameters are shown in Table S1.

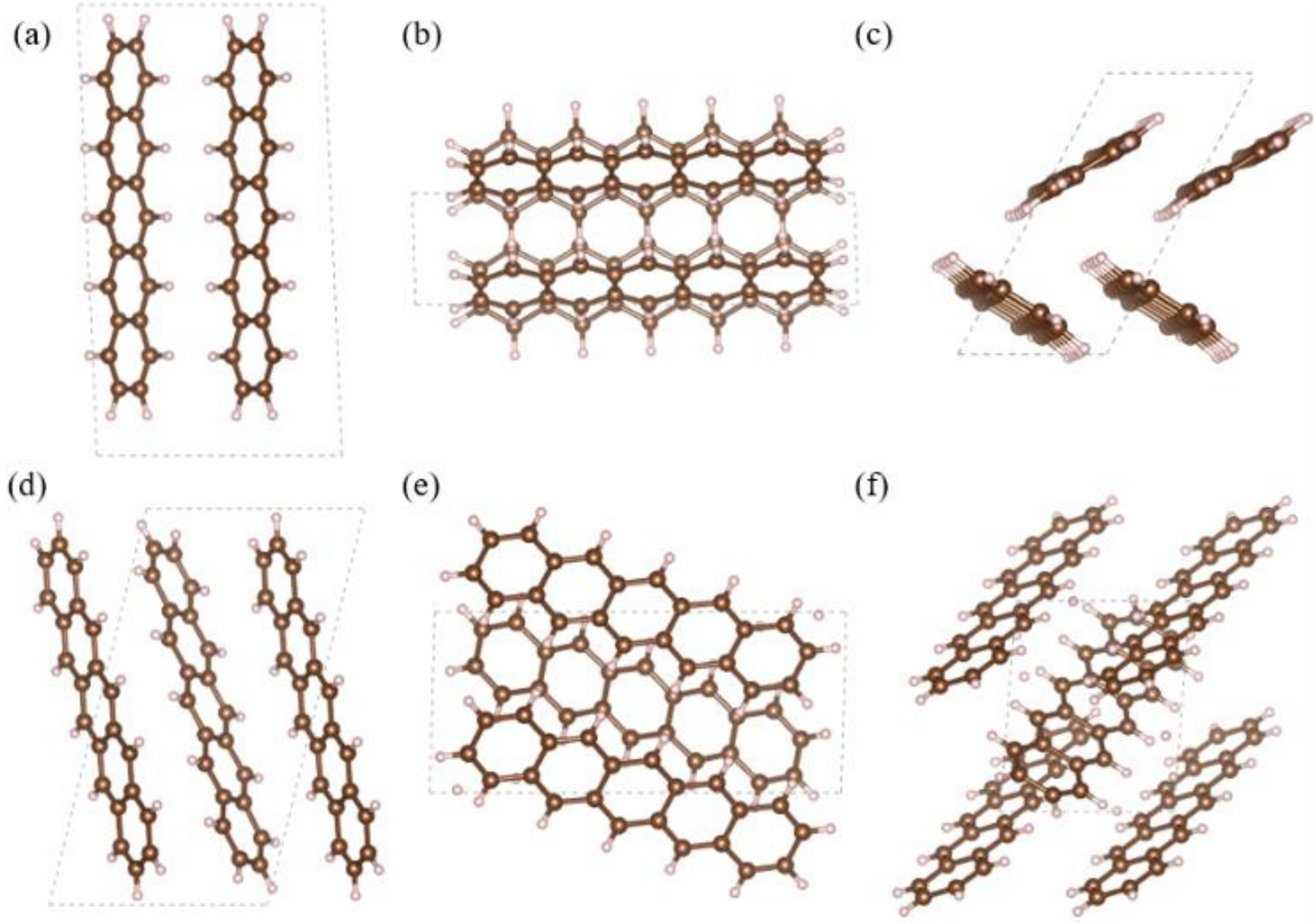

**Figure 2.** Crystal structures of pentacene projected along different crystallographic axes. (a–c) Monolayer phase; (d–f) bulk phase. The projections are taken along the a, b, and c directions from left to right. Carbon and hydrogen atoms are shown in brown and pink, respectively.

To ensure a meaningful comparison with previous results, the monolayer pentacene structure we used is consistent with the structure and lattice parameters employed in the work of Akimov et al[13]. Electronic structure calculations, AIMD simulations and NAC calculations were performed in the CP2K package. The exchange-correlation interactions were described by Perdew-Burke-Ernzerhof (PBE) functional[39] within the generalized gradient approximation (GGA)[40]. Core electrons were represented by Goedecker-Teter-Hutter (GTH) pseudopotentials[41], and the valence electrons were described using the molecularly optimized double-zeta valence polarized Gaussian basis sets (DZVP-MOLOPT-GTH)[42], with the plane-wave cutoff energy set to 450 Ry. To accurately account for van der Waals interactions, Grimme's DFT-D3 empirical dispersion correction was employed[43]. The systems were heated at 300 K for 5 ps in the canonical ensemble, using Nosé-Hoover thermostat[44,45]. Subsequently, AIMD simulations were performed in the microcanonical ensemble with a time step of 1 fs, generating a trajectory of 10 ps.

NAMD simulations were performed using the PYXAID package, employing the CPA-FSSH algorithm to describe nonadiabatic transitions between electronic states. The active space is set to 7 bands (VBM, CBM, CBM+1, CBM+2, CBM+3, CBM+4, CBM+5), where the VBM is the valence band maximum and CBM is the conduction band minimum. Considering single-electron excitation starting from the system at the ground state (GS), for an electronic state that the electron at VBM is excited to CBM, CBM+1, etc., respectively, denoted as SE1, SE2, etc. 200 initial configurations were sampled from the first 2 ps AIMD trajectories, and electrons are initial excited to the highest excited state SE6. For each initial configuration, 1000 independent surface-hopping trajectories are generated and propagated for 3000 fs, corresponding to the trajectories from 2 ps to 5 ps in AIMD. This procedure enables a systematic evaluation of the hot electron relaxation dynamics in crystalline pentacene.

In the following, we present the results of hot electron relaxation dynamics obtained without and with phase correction for comparison. Without phase correction, the populations of hot electrons relaxed from SE6 are shown in Figure 3a, clearly identifying distinct timescales in hot electron relaxation process. In the initial stage, within approximately 500 fs, the system undergoes an ultrafast relaxation where electrons rapidly decay from SE6 to lower-excited states (mainly SE5, SE4, and SE3). Subsequently, the electrons remained at SE1 for a long time before final recombination. This phenomenon is in good agreement with the computational results reported by Akimov et al[13]. Next, the phase consistency of the wavefunctions was taken into account in NAC calculations. We reproduced the NAMD simulations of pentacene using the phase-corrected NAC, and the results are shown in Figure 3b. Electrons on SE6 rapidly relax to lower-excited state within approximately 250 fs. The population of SE1 reaches maximum around 700 fs, after which electrons start transitioning to the ground state. It can be observed that the hot electron relaxation time obtained after introducing phase correction is consistent in scale with that without phase correction, but the relaxation process accelerates to a few ps. Therefore, the phase consistency resulting from phase correction has a significant impact on the hot electron relaxation process in pentacene.

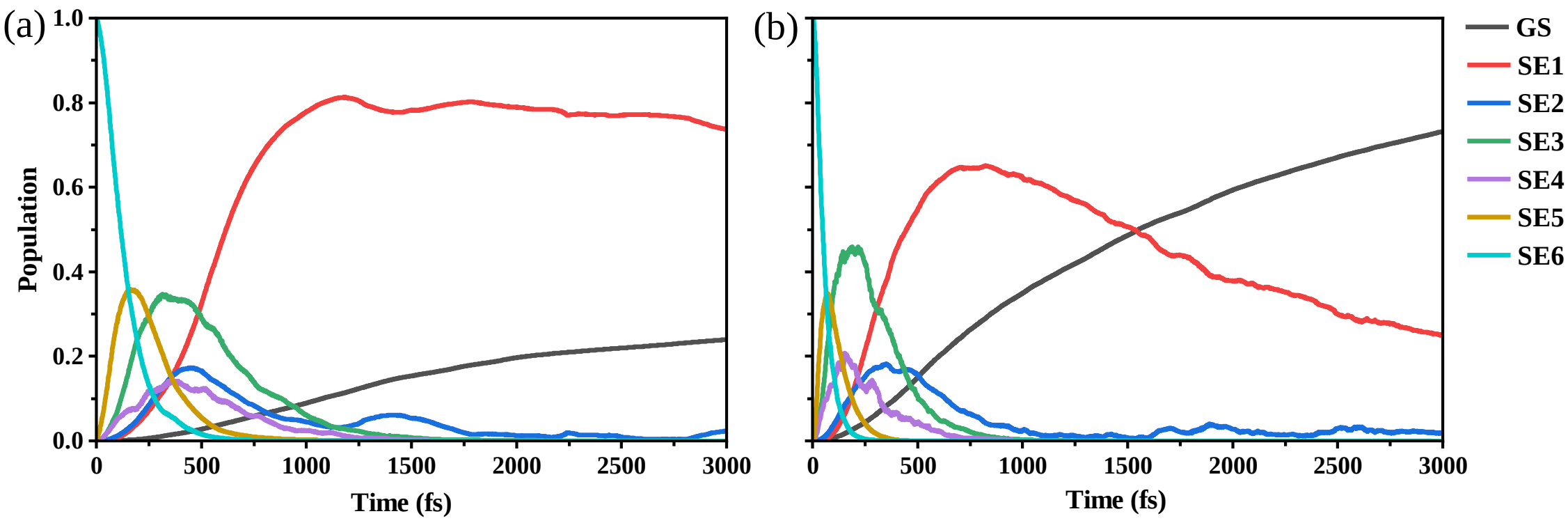

**Figure 3.** Population of GS–SE6 during hot electron relaxation in crystalline pentacene (a) monolayer pentacene without phase correction (b) monolayer pentacene with phase correction. In the figure, GS denotes the ground state (VBM), while SE1–SE6 represent the first to sixth excited states, with SE1 corresponding to the CBM.

Then, we investigated the hot electron relaxation process in bulk pentacene with 288 atoms to explore the applicability and efficiency of our implementation in large supercell. For bulk pentacene, a 2×2×1 supercell was built based on the unit cell. Since the system size is four times larger than the unit cell, the active space for NAMD simulation expands from 7 bands to 25 bands, including 1 ground state and 24 excited states (corresponding to 6 excited states in unit cell). The results of NAMD simulation are shown in Figure 4. Similar to the monolayer phase, the excited electrons initially undergo an ultrafast relaxation within 200 fs, as observed by the fast decay of SE24, corresponding to SE6 in the unit cell. However, there are notable differences in the intermediate and lower-excited states. In particular, both the magnitude of population and relaxation of higher-excited states are extended (such as SE9, corresponding to SE3 in the unit cell), indicating a slower decay of these states compared to the unit cell. Moreover, the long-lived population of SE1 reveals prolonged lifetime of electron at CBM.

The differences between the unit-cell and supercell results can be primarily attributed to following factors. First, the use of a supercell in molecular dynamics simulations significantly reduces finite-size effects present in unit-cell calculations, thereby providing a more accurate description of long-range structural fluctuations and atomic interactions. Second, the supercell introduces band folding in reciprocal space. When performing the calculation using Gamma-only k-point sampling, supercell can effectively capture additional information compared to the unit cell. Based on these two aspects, NAC evaluated in the supercell is expected to be more physical. Finally, the supercell approach enables the description of systems with lower carrier concentrations, which is suggested to be closer to realistic conditions.

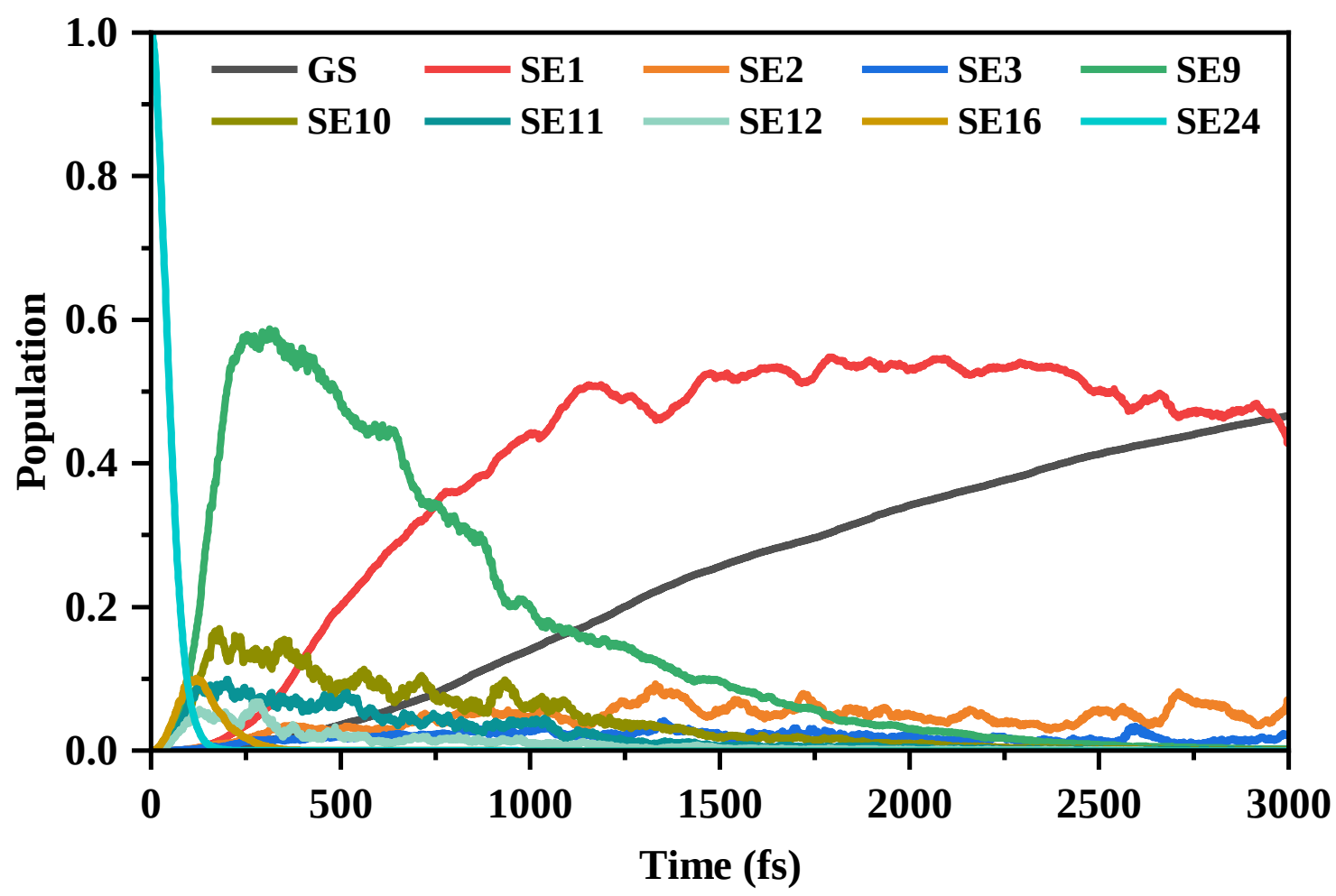

**Figure 4.** Population of GS, SE1–SE24 during hot electron relaxation in 2×2×1 supercell of bulk pentacene. Among the 24 excited states we calculated, SE3–SE8 are represented by SE3, and SE13–SE23 are represented by SE16, while other states with similar characteristics are omitted for clarity.

## 4 Conclusion

In this work, we have developed and implemented an efficient computational framework for evaluating nonadiabatic coupling within the CP2K framework. This approach enables rapid on-the-fly NAC evaluation during AIMD while avoiding costly I/O operations required in conventional interface-based workflows. Combined with the efficiency of the GPW method, it enables NAMD simulations of large-scale systems at significantly reduced computational cost. To validate the reliability of the implementation, we investigated the hot electron relaxation dynamics in crystalline monolayer pentacene, obtaining results that show good agreement with previously reported theoretical studies. The introduction of phase correction ensures the stability of the NAC values, and the results of NAMD simulation demonstrate that phase correction has a pronounced impact on the relaxation process, highlighting its essential role in NAC calculation. Furthermore, there are noticeable differences in the relaxation dynamics between the supercell and unit cell, the higher-excited states persist for a longer duration, and electrons at CBM also have a longer lifetime in the supercell. Overall, this work establishes a time-saving, cost-effective, tech-credible method for NAC evaluation with CP2K framework.

**Acknowledgement**

This work was financially supported by the National Key Research and Development Program of China (No. 2022YFB4200501). The work was supported by the National Natural Science Foundation of China (No. 52225308, No. 52533010), and the Fundamental Research Funds for the Central Universities (JKF-20240722). This research was supported by the high-performance computing (HPC) resources at Beihang University.

# Supporting Information for:

# An Efficient On-the-Fly Nonadiabatic Coupling Framework Integrated into CP2K

Xiaoke He[1,†], Hsiao-Yi Tsai[1,†], Ziwei Chai[2], JunWen Yin[3,*], Zhongfei Xu[4,*], and Li-Min Liu[1,*]

1 School of Physics, Beihang University, Beijing 100191, China

2 School of Physical Science, Great Bay University, Dongguan, 523000, China

3 Scientific Computing Department, Science and Technology Facilities Council, UK Research and Innovation, Daresbury Laboratory, Keckwick Lane, Daresbury WA4 4AD, United Kingdom

4 College of Environmental Science and Engineering, North China Electric Power University, Beijing 102206, China

*Corresponding authors:

Li-Min Liu, Email: liminliu@buaa.edu.cn; Zhongfei Xu, Email: xuzhongfei@ncepu.edu.cn; JunWen Yin, Email: junwen.yin@stfc.ac.uk

†These authors contributed equally to this work

## 1 Fewest Switches Surface Hopping (FSSH) Algorithm

Trajectory Surface Hopping (TSH) is a prominent semiclassical nonadiabatic dynamics method in which the atomic nuclei evolve classically on a single potential energy surface, while nonadiabatic transitions are described by instantaneous hops between different electronic states.[1,2] Therefore, a central issue in TSH is how to define the hopping probability that determines when a hopping between electronic states occurs.

Among the various surface hopping schemes, the FSSH algorithm proposed by Tully has emerged as the standard formulation due to its higher computational efficiency and reduced cost.[3,4] To connect the quantum amplitude to the classical trajectory swarm, the electronic density matrix elements are defined as $\rho_{kl}(t) = a_k(t)a_l(t)^*$, the diagonal element $\rho_{kk}$ and off-diagonal element $\rho_{kl}$ of the density matrix represent the occupation and coherence of electronic states, respectively. Therefore, the TDSE equation can be formulated as the evolution equation of density matrix elements:

$$i\hbar\frac{d\rho_k}{dt} = \sum_l [\rho_{lk}\left(H_{el,kl} - i\hbar\frac{d\mathbf{X}}{dt}\cdot\mathbf{A}_{kl}\right) - \rho_{kl}\left(H_{el,lk} - i\hbar\frac{d\mathbf{X}}{dt}\cdot\mathbf{A}_{kl}\right)] \tag{1}$$

The time-dependent population of electronic state $k$ is given by the sum of population transfers from all other states into state $k$ within a unit time interval, which can be written as

$$\frac{d\rho_{kk}}{dt} = \sum_{l\neq k} b_{kl} \tag{2}$$

where $b_{kl}$ represents the rate of transition from state *l* to state *k*, expressed as

$$b_{kl} = \frac{2}{\hbar}\Im\{\rho_{kl}^* H_{el,kl}\} - 2\Re\left\{\rho_{kl}^*\frac{d\mathbf{X}}{dt}\mathbf{A}_{kl}\right\} \tag{3}$$

In the adiabatic representation, the Hamiltonian matrix is diagonal ($H_{kl} = E_k\delta_{kl}$), and the transition rate is driven purely by the NAC term:

$$b_{kl} = -2\Re\left\{\rho_{kl}^*\frac{d\mathbf{X}}{dt}\mathbf{A}_{kl}\right\} \tag{4}$$

The FSSH algorithm insists that the fractional number of classical trajectories hopping from current state $k$ to state $l$ within a time step $\Delta t$ should exactly match the quantum population flux into state $l$, and a hop should only be considered if the population of state $k$ is actively decaying into state $l$. Therefore, the transition probability $g_{kl}$ is defined as:

$$g_{kl} = \max(0, \frac{b_{lk}\Delta t}{\rho_{kk}}) \tag{5}$$

To satisfy *fewest switches* criterion, if $b_{lk} < 0$, which means the quantum flux is actually flowing from state $l$ back to state $k$, an active hop from state $k$ to state $l$ should not be triggered; therefore, the transition probability $g_{kl}$ is strictly set to 0.

In the original FSSH formulation, energy conservation is enforced through velocity rescaling along the NAC vectors, accompanied by the rejection of frustrated hops. However, in our implementation based on the Classical Path Approximation (CPA), the nuclear trajectories are predetermined. To properly reflect the detailed balance condition during the nonadiabatic dynamics, the velocity rescaling mechanisms are replaced by scaling the original FSSH transition probabilities $g_{kl}$ with a thermodynamic Boltzmann factor $f_{kl}$.[5]

Therefore, the corrected transition probability $\tilde{g}_{kl}$ from state $k$ to state $l$ is evaluated as:

$$\tilde{g}_{kl} = g_{kl} f_{kl} \tag{6}$$

where the Boltzmann factor $f_{kl}$ is defined according to the instantaneous adiabatic energy gap between the target and current states:

$$f_{kl} = \begin{cases} \exp\left(-\frac{E_l - E_k}{k_B T}\right), & if \;\; E_l > E_k \\ 1, & if \;\; E_l \leq E_k \end{cases} \tag{7}$$

Here, $k_B$ is the Boltzmann constant and $T$ is the simulation temperature. This scaling strategy exponentially suppresses the probability of upward transitions, while keeping the downward transition rates unaffected. This Boltzmann scaling ensures the detailed balance during nonadiabatic dynamics.

To determine whether a quantum transition occurs, the FSSH method introduces a uniform random number $\xi \in (0,1)$ to compare with $\tilde{g}_{kl}$. When the random number satisfies

$$\sum_{m=1}^{l-1} \tilde{g}_{km} < \xi \leq \sum_{m=1}^{l} \tilde{g}_{km} \tag{8}$$

it is considered that the electronic state has transitioned from state $k$ to state $l$.

## 2 Lattice parameters of monolayer pentacene and bulk pentacene

**Table 1.** Lattice parameters of monolayer pentacene and bulk pentacene

| | **a** | **b** | **c** | $\boldsymbol{\alpha}$ | $\boldsymbol{\beta}$ | $\boldsymbol{\gamma}$ |
|---|---|---|---|---|---|---|
| **monolayer** | 4.615 Å | 9.904 Å | 16.134 Å | 93.6 ° | 93.2 ° | 62.1 ° |
| **bulk** | 6.275 Å | 7.888 Å | 14.710 Å | 76.0 ° | 87.2 ° | 85.0 ° |